\documentclass[aps,prb,reprint,showkeys,twocolumn]{revtex4-2}

\usepackage{graphicx}
\usepackage[breaklinks=true,colorlinks=true,urlcolor=blue,
citecolor=blue,linkcolor=blue,bookmarks=true,bookmarksopen=true]{hyperref}
\usepackage{bookmark}
\usepackage{tensor}

\begin{document}
	\title{ Symmetry induced phonon renormalization in few layers of 2H-MoTe$ _{2} $ transistors: Raman and first-principles studies } 
	\author{Subhadip Das$^1$, Koyendrila Debnath$ ^{2} $, Biswanath Chakraborty$^{1,3}$, Anjali Singh$^{4}$, Shivani Grover$ ^{2} $, D. V. S. Muthu$^1$, U. V. Waghmare$^2$ and A. K. Sood$^1$}
	
	
	\email{asood@iisc.ac.in}
	\affiliation{$^1$Department of Physics, Indian Institute of Science, Bangalore 560012, India
		\\
		$^{2}$  Theoretical Sciences Unit, Jawaharlal Nehru Centre for Advanced Scientific Research, Bangalore-560064, India \linebreak $ ^{3} $ Present address: Department of Physics, Indian Institute of Technology Jammu, Jammu-181221, J\&K, India\\
		$ ^{4} $ Center for Study of Science, Technology \& Policy (CSTEP), Bangalore 560094, India.}

	\begin{abstract}
		Understanding of electron-phonon coupling (EPC) in two dimensional  (2D) materials manifesting as phonon renormalization is essential to their possible applications in nanoelectronics. Here we report \textit{in-situ} Raman measurements of  electrochemically top-gated 2, 3 and 7 layered  2H-MoTe$ _{2} $ channel based field-effect transistors (FETs). While the E$ ^{1}_{2g} $ and B$ _{2g} $ phonon modes exhibit frequency softening and linewidth broadening with hole doping concentration (\textit{p}) up to $\sim$ 2.3 $\times$10$ ^{13} $/cm$ ^{2} $,  A$ _{1g}$ shows relatively small frequency hardening and linewidth sharpening. The dependence of frequency renormalization of the E$ ^{1}_{2g} $ mode on the number of layers in these 2D crystals confirms that hole doping occurs primarily  in the top two layers, in agreement with recent predictions. We present  first-principles density functional theory (DFT) analysis of bilayer MoTe$ _{2} $ that qualitatively captures our observations, and explain that a relatively stronger coupling of holes with E$ ^{1}_{2g} $ or B$ _{2g} $ modes as compared with the A$ _{1g} $ mode originates from the in-plane orbital character and symmetry of the states at valence band maximum (VBM). The contrast between the manifestation of EPC in monolayer MoS$ _{2} $ and those observed here in a few-layered MoTe$ _{2} $ demonstrates the role of the symmetry of phonons and electronic states in determining the EPC in these isostructural systems.
		
	\end{abstract}

	\maketitle
	\section{Introduction}
\begin{figure*}[t!]
		\includegraphics[width=0.7\textwidth ]{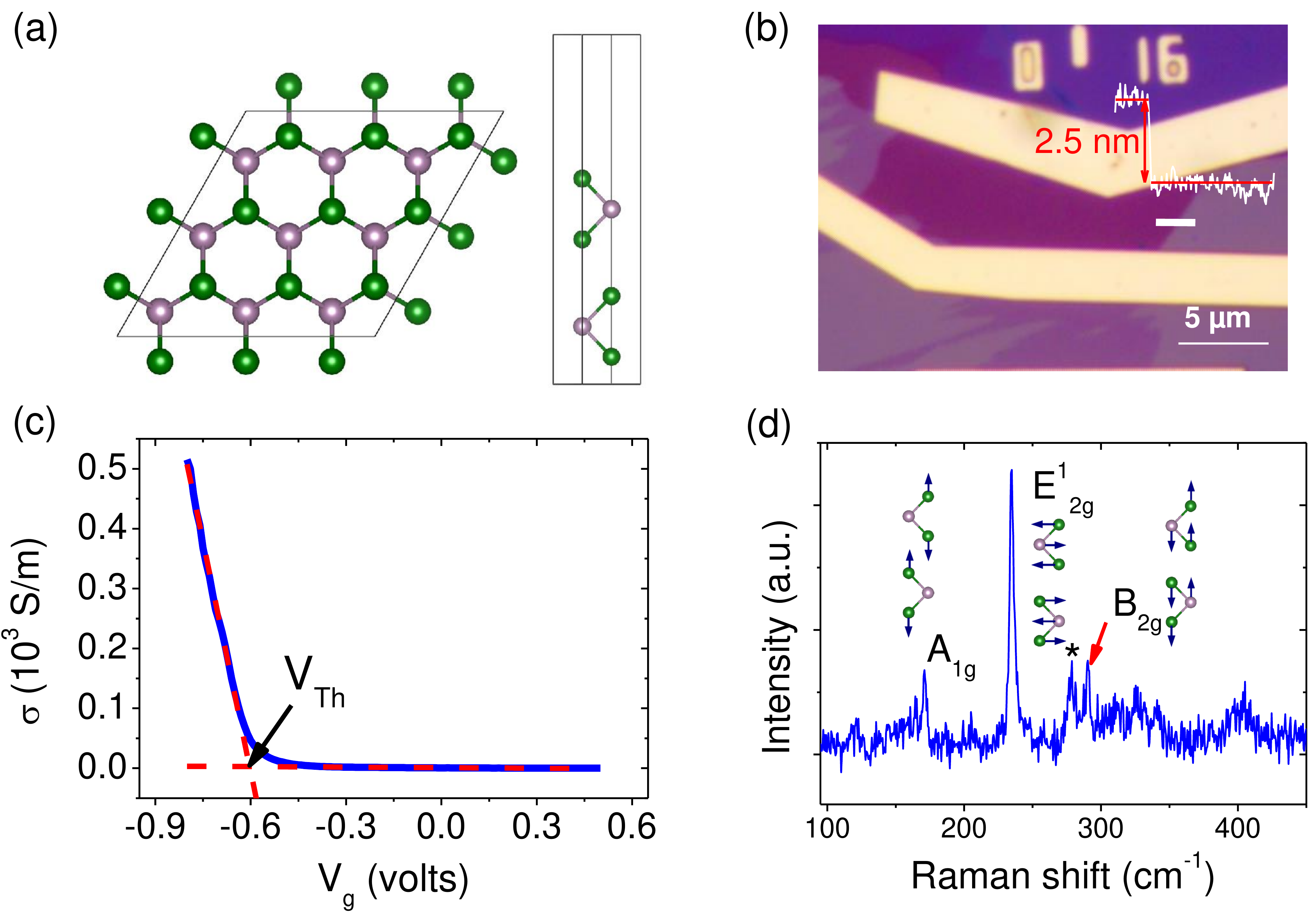}
		
		\caption{(a) Top and side view  of the crystal structure of bilayer 2H-MoTe$_{2}$. Violet and green spheres denote Mo and Te atoms respectively. (b) Optical image of the field-effect transistor (FET) device. Inset image shows the AFM height profile of the nanocrystal.  (c) Transfer characteristics of the trilayer FET device. Red dashed lines show the linear fit to the transistor off and on state. The intersection of these lines indicate the current threshold voltage (V$ _{Th} $).  (d) Raman spectrum  of the trilayer nanocrystal. The star symbol indicates Raman peak from the ionic liquid. Atomic displacements of the phonon modes \cite{doi:10.1021/acs.nanolett.5b02683} are indicated in the figure.}
		
		\label{1}
	\end{figure*}

	The discovery  of unique and remarkable properties of graphene has sparked unprecedented interest in other classes of two dimensional (2D) materials like transition metal dichalcogenides (TMDs, MX$ _{2} $, where M= transition metals (Mo, W, Ti, Nb, Ta) and X= chalcogens (S, Se, Te)) for their potential applications in nano and opto-electronics \cite{wang2012electronics}. Optical and electrical properties of these TMDs  can be easily manipulated by both changing the layer number and carrier doping.  MoTe$ _{2} $ is a member of the group-VI TMD family that crystallizes into three stable phases: Hexagonal ($\alpha$ or 2H) \cite{Puotinen:a03172}, monoclinic ($\beta$ or 1T$ ^{\prime} $) \cite{Brown:a04998} and orthorhombic ($\gamma$ or T$ _{d} $) \cite{qi2016superconductivity}. The 2H phase is semiconducting \cite{doi:10.1021/nl502557g}, whereas the 1T$ ^{\prime} $ phase is a narrow band gap semiconductor \cite{keum2015bandgap}. Similar to other group-VI dichalcogenides, 2H-MoTe$ _{2} $ has a trigonal-prismatic coordinated crystal structure \cite{PhysRevLett.25.362}, consisting of weakly coupled sandwich layers of Te-Mo-Te units, where  Mo-atom layer is enclosed between two Te layers (Fig. \ref{1}(a)) \cite{Puotinen:a03172}. Unlike other TMDs, energy  difference between the 2H and 1T$ ' $ phase is very small ($\sim$  31 meV per formula unit \cite{li2016structural}). This enables easy tuning of the two phases by  strain \cite{doi:10.1021/acs.nanolett.5b03481,hou2019strain}, laser irradiation \cite{Cho625,C8NR06115G} and electron doping \cite{wang2017structural,doi:10.1021/acsnano.9b07095}, making this material an ideal candidate for next generation homojunction devices \cite{doi:10.1021/acsnano.9b02785}. From electron doping (n) induced transition from 2H to 1T$ ' $   phase in multilayer MoTe$ _{2} $, Zakhidov \textit{ et al.} recently suggested that doped electrons by ionic liquid (IL) gating  are localized on the top few layers of the nanocrystal \cite{doi:10.1021/acsnano.9b07095}, consistent with previous theoretical calculations \cite{PhysRevB.91.155436}.  
	
	Bulk MoTe$ _{2} $ has an indirect band gap of  $\sim$  1.0 eV \cite{doi:10.1002/pssb.2220940132,Grant_1975} which becomes a direct band gap semiconductor in a monolayer  with an emission peak of excitonic photoluminescence (PL) spectrum in the near-infrared range (  $\sim$  1.1 eV) \cite{doi:10.1021/nl502557g}. This enables the material to be a highly sensitive photodetector \cite{doi:10.1021/acsami.6b14483, doi:10.1063/1.4941996} and light-emitting diode \cite{li2017room,bie2017mote}.  With  device performance at par with its sister compounds MoS$ _{2} $ and MoSe$ _{2} $ \cite{doi:10.1021/nn501013c}, MoTe$ _{2} $ shows ambipolar transport properties \cite{doi:10.1021/acsnano.5b00736}, which has been recently implemented as a p-n homojunction rectifier device with low charge trapping at the junction interface \cite{C8NR10526J}. Since the exploration of these properties has been mostly carried out in monolayer regime, investigation of the charge localization at high gate bias in few layers of the nanocrystal can open up new possibilities in the field of opto-electronics. 
	
	Carrier concentration in a semiconductor can be modulated by injecting resonant photons from light emitting diode \cite{doi:10.1021/acsnano.9b04163,doi:10.1021/acs.jpclett.0c00993,doi:10.1021/acs.jpclett.0c01642}, substitutional doping during growth process \cite{doi:10.1021/acsami.8b08773} and application of an electrostatic field on the channel of a FET \cite{PhysRevLett.84.2941,PhysRevLett.105.256805, PhysRevB.85.161403, Chakraborty_2016}. Electrolyte gating has gained immense interest recently for electrostatic modulation of carrier density up to $\sim 10^{15}$/cm$ ^{2} $ \cite{doi:10.1002/adfm.200801633} owing to their large dielectric capacitance. On application of gate voltage, ions move inside the electrolyte to screen the applied electric field to form few Angstroms \cite{doi:10.1021/nl049937e} thick double layers of ions near the  device and gate electrode \cite{Gebbie7432,Jurado2017}. However electrolyte gating is well known source of electrostatic disorders \cite{PhysRevLett.105.036802,gallagher2015high}  and is best suited for disorder robust systems such as superconductors \cite{Ueno2008,Ye2010}.

	As Raman spectroscopy does not require any sample preparation, it has been extensively used as a non-invasive, contact-less, fast and accurate tool to determine strain \cite{Karki_2020}, doping effects \cite{wang2017structural}, layer number \cite{Grzeszczyk_2016, doi:10.1021/acs.nanolett.5b02683}, crystal orientation \cite{Song2017}, structural transitions between different polytypes \cite{C9NR10383J,Bera_2017,doi:10.1021/acs.nanolett.5b03481,wang2017structural,doi:10.1021/acsnano.9b07095,Cho625,C8NR06115G,PhysRevB.99.024111} in fewlayer MoTe$ _{2} $ devices in ambient as well as different sample environments.  Furthermore, Raman scattering has been employed in various 2D materials to measure electron-phonon coupling (EPC) that governs electronic  transport properties \cite{PhysRevLett.84.2941,PhysRevLett.105.256805}. For n-type semiconducting MoS$ _{2} $, symmetry of the conduction band minimum (CBM) determines EPC of the A$ _{1g} $ and E$ ^{1}_{2g} $ modes \cite{PhysRevB.85.161403}. In ambipolar phosphorene transistor, electrons and holes couple differently to phonons as CBM and valence band maximum (VBM) possess different orbital symmetries \cite{Chakraborty_2016}. Although the electronic band structure of monolayer MoTe$ _{2}$  is similar to MoS$ _{2}$ \cite{PhysRevB.91.155436}, the VBM of the former remains at the K-point from single to three layers \cite{doi:10.1021/nl903868w,PhysRevB.91.155436}. Thus a study of the EPC in few layer MoTe$ _{2} $, an intrinsic p-type semiconductor \cite{doi:10.1021/acsami.0c04339,doi:10.1002/adma.201606433},  will reveal asymmetry of phonon coupling with holes and electrons in these hexagonal polytypes of TMDs.

	Bulk MoTe$ _{2} $ belongs to D$ _{6h} $ point-group \cite{PhysRevB.90.115438} having six Raman active modes (A$ _{1g} $+ 2B$ _{2g} $+ E$ _{1g} $+ 2E$ _{2g} $) \cite{doi:10.1021/acs.nanolett.5b02683}. A$ _{1g} $ and E$ _{2g} $ modes have vibrations perpendicular to and along the basal plane of the lattice, respectively \cite{doi:10.1021/acs.nanolett.5b02683}. The in-plane E$ _{1g} $ mode is absent in backscattering configuration \cite{doi:10.1021/acs.nanolett.5b02683}. Notably,  the translation symmetry along the  \textit{z}-direction is broken in a few layer nanocrystal, reducing the symmetry to  D$ _{3h} $ and D$ _{3d} $ for  odd  and even layers of MoTe$ _{2} $, respectively  \cite{PhysRevB.90.115438}. Thus, the out-of-plane inactive mode B$ _{2g} $ in bulk becomes Raman active in few layers and shows highest intensity in a bilayer nanocrystal \cite{doi:10.1021/nn5007607}. For odd layer nanocrystal, the inversion symmetry breaks, making some modes both Raman and infrared active \cite{doi:10.1021/acs.nanolett.5b02683}. For simplicity, the Raman modes of even and odd layers of MoTe$ _{2} $ in this paper are represented by the bulk phonon symmetry group of equivalent atomic displacements  (see table-S1 of the supplemental material (SM)).

	In the present study,  we measure in-operando  optical phonons in a few layers of 2H-MoTe$ _{2} $ based field-effect transistors (FETs) as a function of hole doping concentration (\textit{p}) up to  $\sim$  2.3$\times$10$ ^{13} $ cm$ ^{-2} $. The modes involving both metal and chalcogen atom vibrations, E$ ^{1}_{2g} $ and B$ _{2g} $ \cite{doi:10.1021/acs.nanolett.5b02683} show phonon softening and linewidth broadening while the  A$ _{1g} $ mode with out-of-plane vibrations of only the chalcogen atoms  \cite{doi:10.1021/acs.nanolett.5b02683}, shows in contrast, relatively smaller phonon hardening and linewidth sharpening. The frequency renormalization comparison of E$ ^{1}_{2g} $ mode from 2, 3 and 7 layer devices indicate that the doping is confined to only two top layers of the  nanocrystal.  We have carried out first-principles density functional theory (DFT) calculations on a bilayer MoTe$ _{2} $ transistor for understanding the experimental results. We show that the holes couple weakly with the A$ _{1g} $ mode as compared to  E$ ^{1}_{2g} $  and B$ _{2g} $ modes and demonstrate that different  orbital symmetries of the VBM and CBM at the K-point of MoTe$ _{2} $ and MoS$ _{2} $, respectively, contribute to their contrasting EPC.

	\section{Results and discussion}
	\subsection{Experimental results}

	Bulk MoTe$ _{2} $ crystals were mechanically exfoliated and transferred to a clean Si/SiO$ _{2} $ (285 nm) substrate. Device contacts were fabricated by first patterning them in electron-beam lithography followed by thermal evaporation of 5 and 50 nm thick chromium and gold respectively. The optical image of the two-probe device is shown in Fig. \ref{1}(b).  Atomic force microscope (AFM) measurement in tapping mode  (inset graph of Fig. \ref{1}(b)) confirms the nanocrystal thickness to be $\sim$ 2.5 nm ($\sim$ 3 monolayers).  A drop of 1-ethyl-3-methylimidazolium bis(trifluoromethylsulfonyl)imide (EMIM-TFSI) IL was drop casted on top of the device channel for electrochemical top gating.  Electrical measurements were done using a Keithley 2400 source meters. Conductivity ($\sigma$) as a function of gate voltage (V$ _{g} $) shows hole transport (Fig. \ref{1}(c)) due to unintentional doping from the environment. Previous experiments done on few layer MoTe$ _{2} $ show formation of Mo-O bonds on tellurium vacancy sites which shift the Fermi level towards the VBM, making MoTe$ _{2} $ intrinsically hole doped \cite{doi:10.1002/adma.201606433,doi:10.1021/acsami.0c04339}. To determine hole doping, \textit{p}, from V$ _{g} $, we use parallel plate capacitor formula, $p= C_G(V_g-V_{Th}) $, where V$ _{Th} $ is the current threshold voltage, and the gate capacitance of the IL, C$ _{G} $ is taken to be   $\sim$  5.9 $\mu$F/cm$ ^{2} $ \cite{doi:10.1063/1.2437663}. Consistent with previous reports \cite{doi:10.1002/adma.201305845,doi:10.1063/1.4901527}, the device shows field-effect mobility  $\sim$  1.85 cm$^{2}$/V.s and current on/off ratio  $\sim$  10$ ^{5} $.
	
	\begin{figure*}[t!]
		\includegraphics[width=0.7\textwidth ]{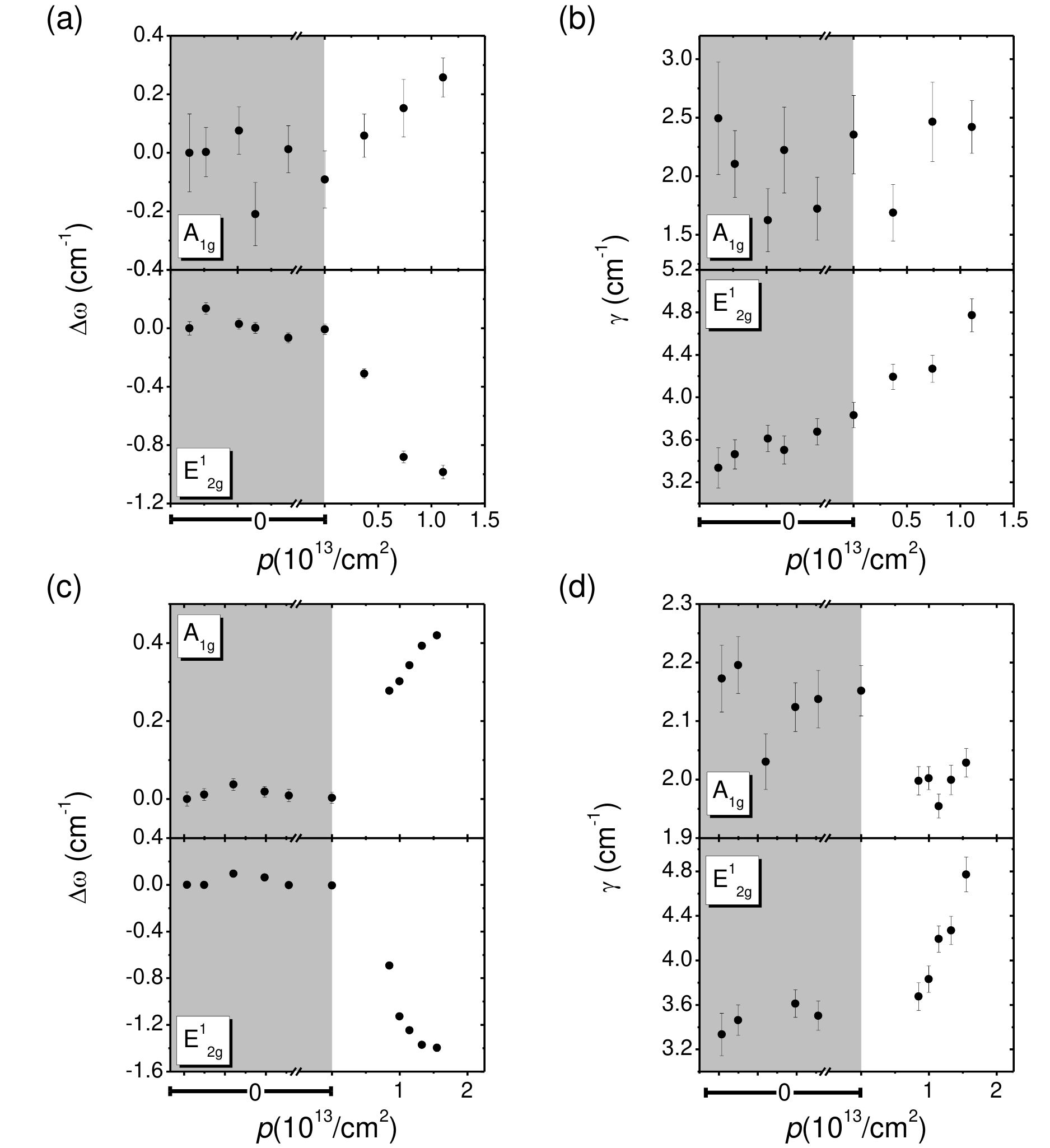}
		
		\caption{ Change in the frequency ($\Delta\omega$=$\omega_{n\neq0}$-$\omega_{n=0}$) and linewidth ($\gamma$) of the Raman modes with hole doping concentration (\textit{p})  for (a and b) trilayer and (c and d) bilayer nanocrystal, respectively. Gray regions represent the zero doped state (V$ _{g}\leq $ V$ _{Th} $). Change in the horizontal axis increments below zero doping is represented by the break symbol. }
		
		\label{2}
	\end{figure*}

	\textit{In-situ} transport and Raman measurements in backscattering configuration were done at room temperature using LabRAM HR-800 Evolution spectrometer having 1800 lines/mm gratings and a Peltier cooled CCD detector. Excitation laser of 532 nm wavelength was focused using a 50$\times$ long working distance objective with incident power less than 0.5 mW to avoid heating. Raman spectrum  of the trilayer nanocrystal is shown in Fig. \ref{1}(d). At each gate voltage, the peaks are fitted with a sum of Lorentzian functions to extract the phonon frequency ($\omega$) and linewidth ($\gamma$). Figs. \ref{2}(a) and (b) show the change in phonon frequency from zero doped state ($\Delta\omega=\omega_{n\neq0}-\omega_{n=0}$) and linewidth  of the trilayer nanocrystal, respectively, with hole doping concentration up to $\sim$ 1.1 $\times$ 10$ ^{13}$/cm$^{2} $. The frequency of the in-plane mode E$ ^{1}_{2g} $  decreases  and linewidth broadens whereas the out-of-plane mode, A$ _{1g} $  shows in contrast, relatively small phonon hardening and linewidth sharpening. We have repeated our experiments  for a bilayer nanocrystal. We have used a FET device with both bilayer and multilayer channels in parallel to confirm our doping effect in the electrical transfer characteristic (see Fig. S1(a) of the SM). Electrochemical gating induced ions sit very close ($\sim$few Angstrom \cite{doi:10.1021/nl049937e}) to the semiconductor surface in a FET. Using EMIM-TFSI gating on a hexagonal boron-nitride enclosed strontium titanate (STO), a two-dimensional electron gas system (2DES), Gallagher \textit{et al.} \cite{gallagher2015high} have shown that disorders induced by the IL reduces the mobility by an order of magnitude.  Xia \textit{et al.} \cite{PhysRevLett.105.036802} theoretically explained this effect in terms of 2D percolative transport from trapped carriers due to the ions induced by the IL at the semiconductor-electrolyte interface. Hence, the mobility suppression will be higher in monolayer and bilayer channels resulting in negligible transistor performance.

	As the intensity of the A$ _{1g} $ Raman mode is weak for 532 nm laser excitation for few layers of MoTe$ _{2} $ \cite{doi:10.1021/nl502557g}, we have used excitation wavelength of 660 nm (Fig. S1(b) of the SM) for this device. With hole doping concentration up to $\sim$ 1.5 $\times$ 10$ ^{13} $/cm$^{2}$, similar to the trilayer nanocrystal (Figs. \ref{2}(a) and (b)),   A$ _{1g} $ mode shows a trend of phonon hardening and linewidth sharpening to a small extent and the E$ ^{1}_{2g} $ mode shows phonon softening and linewidth broadening (Figs. \ref{2}(c) and (d)). The nature of the A$ _{1g} $ mode as reported by Grzeszczyk \textit{et al.} \cite{Grzeszczyk_2016}, depends on the thickness of the MoTe$ _{2} $ nanocrystal. They have shown that at 633 nm of laser excitation, A$ _{1g} $ is a single peak both in monolayer and bilayer nanocrystals \cite{Grzeszczyk_2016}. However, the intensity of the peak drops below E$ ^{1}_{2g}$ and splits into multiple peaks in multilayer nanocrystals of 2H-MoTe$ _{2} $  \cite{Grzeszczyk_2016}. The single peak of the A$ _{1g} $ mode in the observed Raman spectrum at 660nm of laser excitation shown in Fig. S1(b) combined with AFM data in Fig. S1(a), confirm that the doping dependence presented in Figs. \ref{2}(c) and (d) are indeed from the bilayer part of the device (Fig. S1(a)). In addition, the observed trend of $\Delta\omega$ and $\gamma$  is similar to the isolated trilayer nanocrystal (Figs. \ref{2}(a) and (b)), confirming the effect of hole doping. However, the transfer characteristics shown in Fig. S1(a) is dominated by the multilayer nanocrystal. Furthermore, the threshold voltage for Raman frequency and linewidth shift of the bilayer nanocrystal (Figs. \ref{2}(c) and (d)) matches well with the threshold voltage in the transfer characteristics (Fig. S1(a)).  It is to be noted that for similar reasons, parallel channels of bulk and monolayer of black phosphorus nanocrystal were used to determine phonon renormalisation with doping \cite{Chakraborty_2016}.
	
	\begin{figure*}[t!]
		\includegraphics[width=0.7\textwidth ]{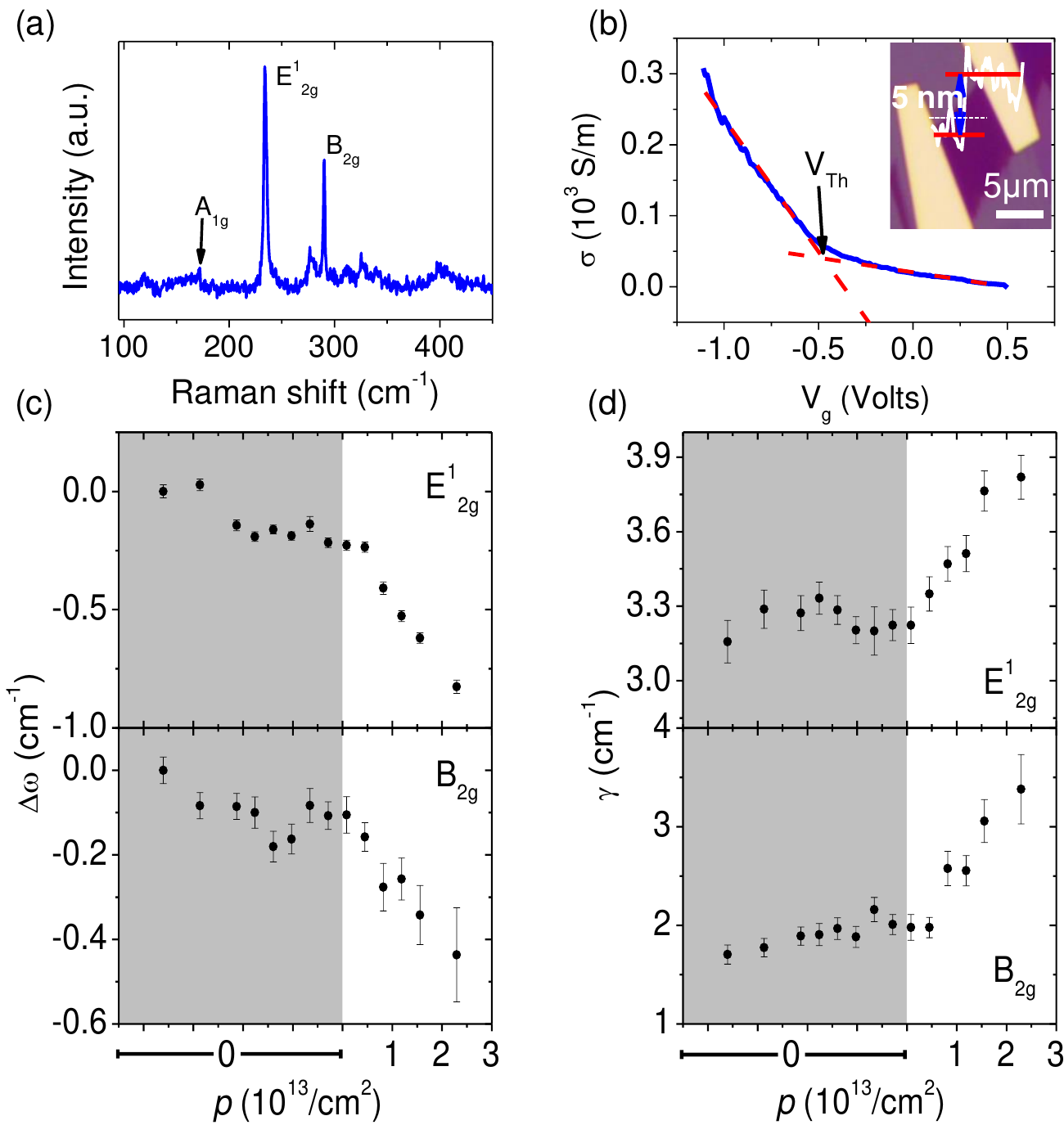}
		
		\caption{(a) Raman spectrum of the seven-layer nanocrystal. (b) Transfer characteristics of the FET device. Similar to the fit on Fig. \ref{1}(c), V$ _{Th} $ is indicated in the figure. Inset graph  shows the AFM height profile of the nanocrystal. (c) $\Delta\omega$ and (d) $\gamma$ versus \textit{p} of  E$ ^{1}_{2g} $ and B$ _{2g} $ modes. }
		
		\label{3}
	\end{figure*}

	As the frequency of the Raman mode from the IL (see Fig. S2 of the SM ) is close to the B$ _{2g} $  mode at $\sim$ 290 cm$ ^{-1} $ (Fig. \ref{1}(d)), we did experiments on a seven-layer thick nanocrystal, where the Raman signal (with 532 nm wavelength of laser excitation) is more prominent (Fig. \ref{3}(a)). Fig. \ref{3}(b) shows the thickness of the nanocrystal to be  $\sim$ 5 nm ($\sim$ seven layers) from AFM measurement (inset graph of Fig. \ref{3}(b)).  The device transfer characteristics in Fig. \ref{3}(b) shows hole field-effect mobility of  $\sim$  0.41 cm$ ^{2} $/V.s and current on/off ratio  $\sim$  10$^{2} $.  Using SiO$ _{2} $ back gate, Pradhan \textit{et al.} \cite{doi:10.1021/nn501013c}, showed similar transistor characteristic of trilayer and seven-layer nanocrystals, since few-layer MoTe$ _{2} $ is an indirect bandgap semiconductor with small change in bandgap from monolayer to bulk \cite{doi:10.1021/nl502557g}. At ambient condition, the seven-layer nanocrystal is initially unintentionally hole doped (Fig. \ref{3}(b)) as compared to the trilayer nanocrystal (Fig. \ref{1}(c)). Thus the dielectric screening of the gate voltage can account for the relatively smaller mobility and current on-off ratio in seven layer crystal as compared with the trilayer nanocrystal. With hole doping up to $\sim2.3\times10 ^{13} /cm ^{2} $, the in-plane vibrational mode, E$ ^{1}_{2g} $ shows  phonon softening and linewidth broadening (Figs. \ref{3}(c) and (d)), similar to trilayer and bilayer nanocrystal (Fig. \ref{2}).  The B$ _{2g} $ mode, although having similar vibrational displacements to  the  A$ _{1g} $ mode \cite{doi:10.1021/acs.nanolett.5b02683}, shows phonon renormalization as the in-plane E$ ^{1}_{2g} $  mode.
	
	\begin{figure*}[t!]
		\includegraphics[width=0.8\textwidth ]{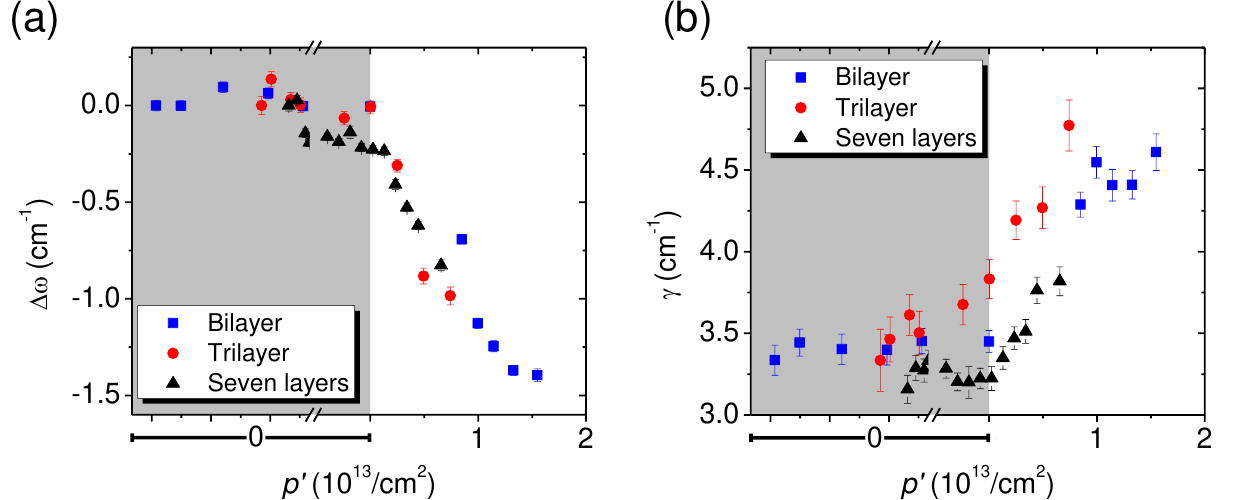}
		
		\caption{(a) $\Delta\omega$ and (b) $\gamma$  with average doping concentration ($ p^{\prime} $) of the E$ ^{1}_{2g} $ mode  in 2, 3 and 7 layer  nanocrystal.}
		
		\label{4}
	\end{figure*}

	Brumme \textit{et al.}   \cite{PhysRevB.91.155436} theoretically showed that the charges localize only in the topmost two layers due to screening effects in MoTe$ _{2} $ based FET devices. From electron doping induced 2H to 1T$ ^{\prime} $ phase transition from bulk to  monolayer MoTe$ _{2} $, Zakhidov \textit{et al.} experimentally showed that the gating by IL causes electrons to be confined in the topmost few layers \cite{doi:10.1021/acsnano.9b07095}. Consistent with these reports, we observe a smaller change in $\Delta\omega$ and $\gamma$ of E$ ^{1}_{2g} $ mode for a given gate voltage as the layer number increases (Figs. \ref{2}, \ref{3}(c) and \ref{3}(d)). Taking the applied doping (\textit{p}) to be limited to the top two layers rather than the entire nanocrystal, the average doping ($p^{\prime}$) for a N-layer nanocrystal is scaled as $p^{'}=p\times 2/N$. With $ p^{'} $, $\Delta\omega$ of  E$ ^{1}_{2g} $ from  2, 3 and 7 layer nanocrystal scale very well with each other (Fig. \ref{4}(a)), justifying that the doping is confined  to top two layers of the nanocrystal. The scaling of $\gamma$ for the three devices (Fig. \ref{4}(b)) is modest with $ p' $.

	\subsection{Theoretical analysis}
	
	\begin{figure*}[t!]
		\includegraphics[width=0.7\textwidth ]{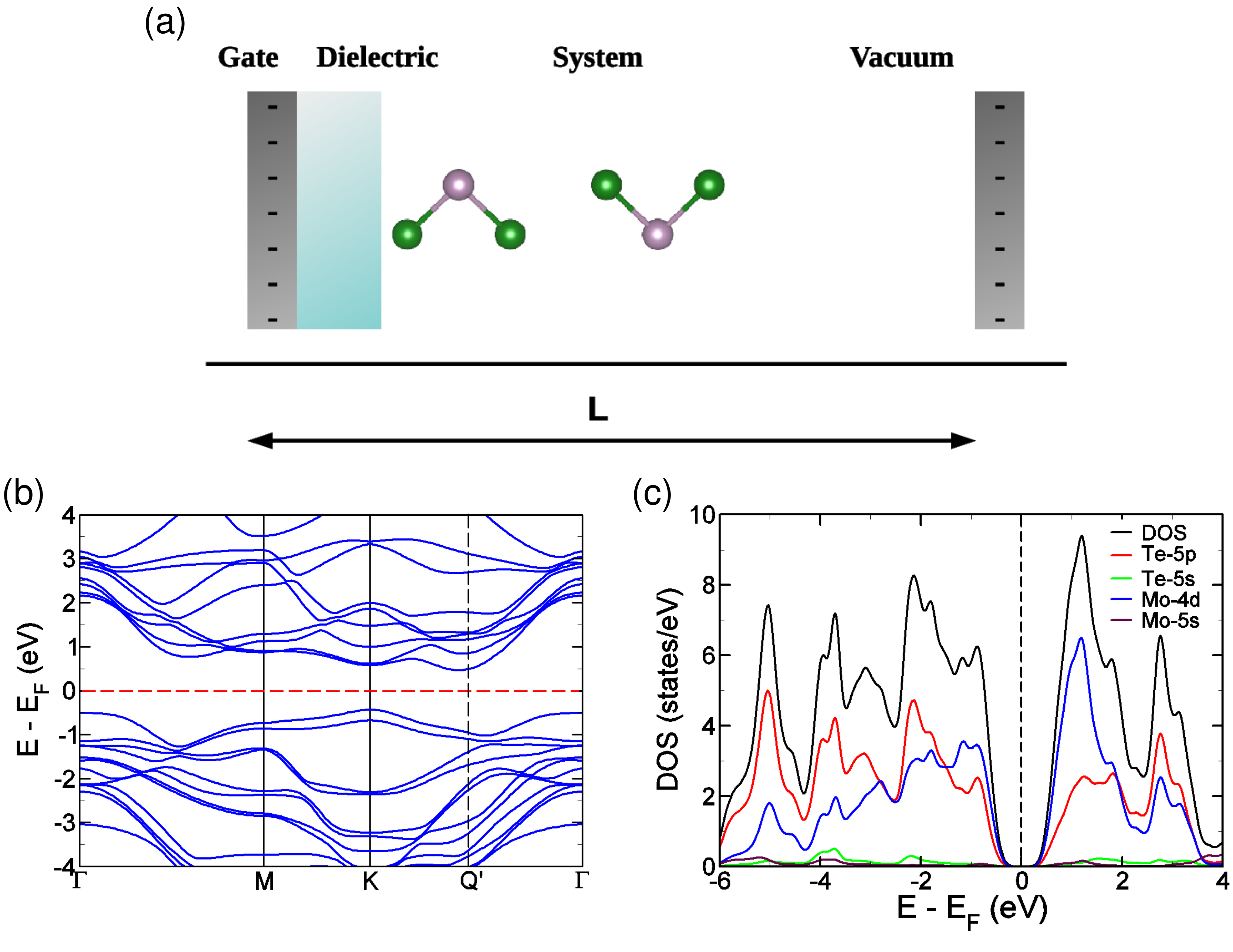}
		
		\caption{(a) Schematic illustration of an FET setup simulated in a periodically repeated unit cell where the layers of 2H-MoTe$_{2}$ is placed in front of a charged plane mimicking the metallic gate (shown with gray color plate). The layers are doped with holes, such that the charged plane is charged with the same magnitude of opposite charges. To mimic the dielectric separation layer, we include a potential barrier (shown in blue). The length of the unit cell along the \textit{z}-direction is given by \textit{L}. (b) Electronic structure of bilayer 2H-MoTe$_{2}$ calculated including the effect of spin-orbit coupling, shows it to be an indirect band gap semiconductor with VBM at K and CBM at Q\textsuperscript{'} point (Q\textsuperscript{'} point is along $\Gamma$-K direction) with a band gap of 0.88 eV. (c) Projected density of states of bilayer 2H-MoTe$_{2}$ shows a  strong coupling between the Mo d orbitals and Te p orbitals  evident in their joint contributions to states near the gap.}
		
		\label{5}
	\end{figure*}
	
	Our first-principles DFT calculations of the bilayer were carried out with Quantum ESPRESSO (QE)  package \cite{giannozzi2009quantum}, in which we treat only the valence electrons by effectively replacing the potential of ionic cores with pseudopotentials. Exchange-correlation energy of electrons is included within a generalized gradient approximation (GGA) \cite{hua1997generalized} in the functional form parametrized by Perdew, Burke, and Ernzerhof \cite{perdew1998perdew}. We include spin-orbit coupling (SOC) through use of relativistic pseudopotentials and a second variational procedure \cite{PhysRevB.71.115106}. Kohn-Sham wave functions and charge density were represented in plane wave basis sets truncated at energy cut-offs of 40 Ry and 320 Ry respectively. A vacuum layer of 10 {\AA} has been introduced parallel to MoTe$_{2}$ layer (perpendicular to \textit{z}-direction) to weaken the interaction between the layer and its periodic images. Brillouin zone (BZ) integrations were sampled on uniform $24\times24\times1$ mesh of \textbf{k}-points. The discontinuity in occupation numbers of electronic states was smeared using a Fermi-Dirac distribution function with broadening temperature of \textit{k}\textsubscript{B}T = 0.003 Ry. We include van der Waals (vdW) interaction using PBE + D2 parametrized scheme of Grimme \cite{grimme2004accurate}.

	\begin{figure}[t!]
		\includegraphics[width=\columnwidth ]{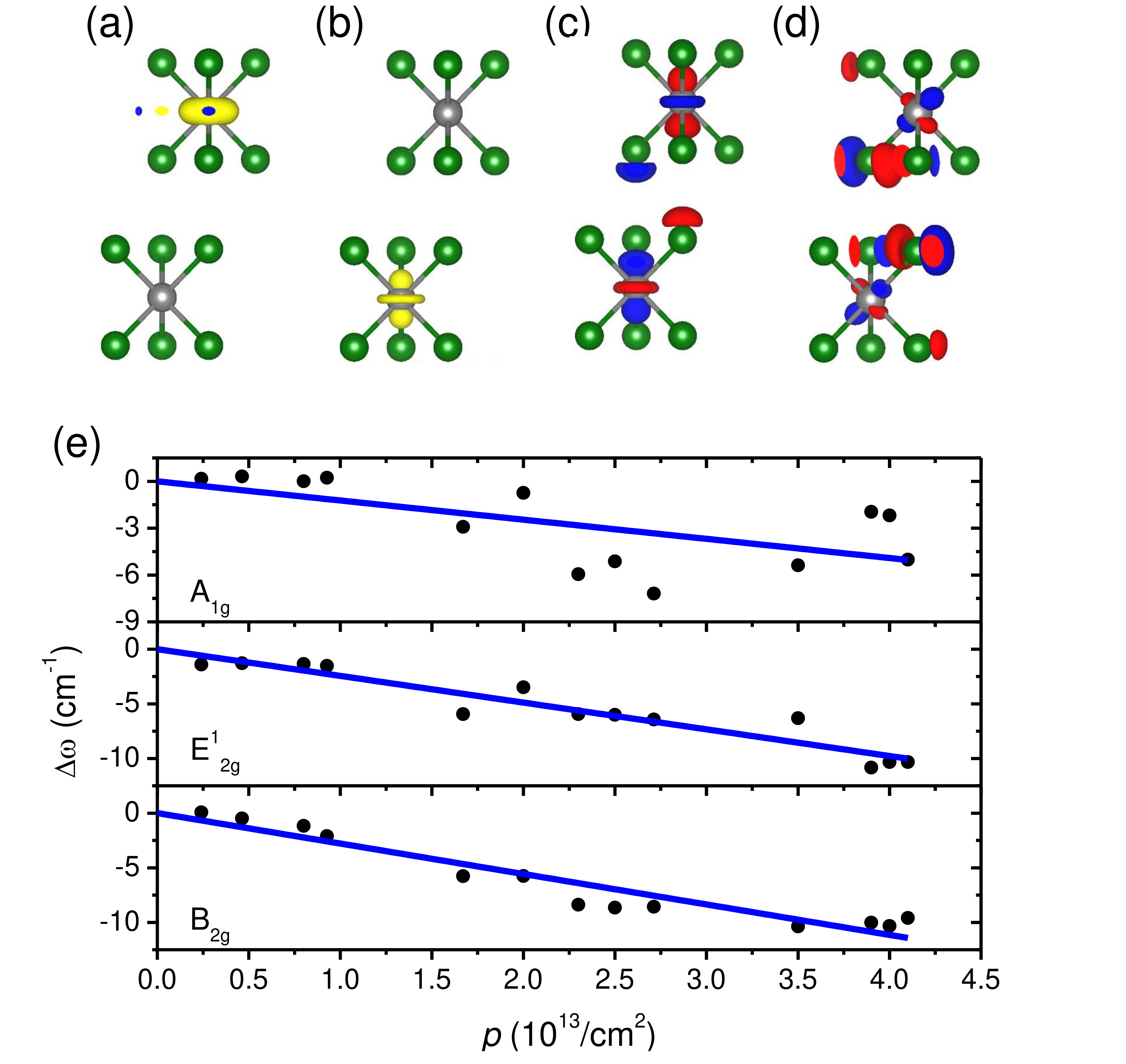}
		
		\caption{ Isosurfaces of wavefunctions of states at (a) VBM  and (b) CBM at K-point showing in-plane d$ _{xy} $ and out-of-plane d$ _{z^2} $ orbital (Mo) character respectively. Isosurfaces of wavefunctions of states at (c) VBM and (d) CBM at $\Gamma$-point. Clearly, VBM at $\Gamma$-point is formed of out-of-plane (Mo d$ _{z^2} $ and Te p$ _{z} $ ) orbitals and CBM has contributions mainly from out-of-plane Mo and in-plane Te states (Mo d$ _{z^2} $ and Te p$ _{x/y} $). (e) Changes in frequencies of A$_{1g}$, E$ ^{1}_{2g} $ and B$_{2g}$ modes as a function of hole concentration \textit{p}. E$ ^{1}_{2g} $ and B$ _{2g} $ modes soften more significantly with hole doping than the A$ _{1g} $ mode. Blue lines represent linear fit to the data with slope of -1.23, -2.44 and -2.78 cm$ ^{-1} $/(10$ ^{13} $cm$ ^{-2} $) for A$ _{1g} $, E$ ^{1}_{2g} $ and B$ _{2g} $ modes, respectively.}
		
		\label{6}
	\end{figure}
	
	We have used the FET setup \cite{PhysRevB.89.245406} as implemented in QE package to treat gating electric field. A 2D charged plate modeling the gate electrode is placed at \textit{z}= 0.019 \textit{L}. A potential barrier with a height of \textit{V}\textsubscript{0} = 0.09 Ry and a width of \textit{d\textsubscript{b}} = 0.1 \textit{L} is used to model the dielectric layer, preventing ions from moving too close to the gate electrode (Fig. \ref{5}(a)). Dynamical matrices were calculated within the Density Functional Perturbation Theory (DFPT) \cite{baroni2001phonons} on a $3\times3\times1$ mesh of \textit{q}-vectors in the Brillouin zone. Fourier interpolation of these dynamical matrices was done to obtain phonon frequencies at arbitrary wavevectors and dispersion along the high symmetry lines in the Brillouin zone.

	Bilayer 2H-MoTe$ _{2} $ has a hexagonal structure, where covalently bonded layers of Te-Mo-Te atomic planes are stacked along \textit{c}-axis interacting via weak vdW interaction. The periodic unit cell of bilayer 2H-MoTe$ _{2} $ is characterized by a stacking sequence \textit{AbABaB},
	where Wyckoff positions \textit{A}, \textit{B} label basal Te atomic planes and \textit{a},\textit{b} label Mo atomic planes of the hexagonal lattice (Fig. \ref{1}(a)). Our estimate of the lattice parameter \textit{a} (= \textit{b}) is 3.53 {\AA} which matches well with previous experimental value \cite{doi:10.1139/v61-110}. Bilayer 2H-MoTe$_{2}$ is an indirect band gap semiconductor with VBM at K and CBM at Q\textsuperscript{'} point (along $\Gamma$-K direction) separated by a gap of 0.88 eV (Fig. \ref{5}(b)). From the projected electronic density of states (DOS) (calculated without SOC) (Fig. \ref{5}(c)), it is evident that there is a rather strong coupling between the Mo d orbitals and Te p orbitals, contributing to states near the gap. Visualization of wavefunctions of states at VBM  and CBM  at $\Gamma$ and K-points confirms the contribution of specific d-orbitals of Mo and p-orbitals of Te (Fig. \ref{6}(a-d)). The doped holes  occupy the states  at the K-point. With increasing hole-doping, holes continue to accumulate in states at K-point because energy separation between valence band of Q\textsuperscript{'}  valley and valence band states at K-point is rather large ( $\sim$ 490 meV) (Fig. \ref{5}(b)). We find that inclusion of SOC in our calculations leads to notable reduction in the indirect band gap by 60 meV and hence we have included SOC in further calculations.

	We simulated hole doping in 2H-MoTe$_{2}$ bilayer by adding a small fraction of holes (close to the experimental doping concentration) to its unit cell. From the changes in calculated frequencies with their linear fits (Fig. \ref{6}(e)), it can be seen that $\Delta\omega$ for A$ _{1g} $ mode is $\sim-5.1$ cm $^{-1} $  for $p=4.1\times10 ^{13} /cm ^{2} $, in contrast to  corresponding higher softening of E$ ^{1}_{2g} $ mode by 10.2 cm$ ^{-1} $ and B$ _{2g} $ mode by 11.4 cm$ ^{-1} $. The magnitude of the slope $S$ ($ =|d(\Delta\omega)/dp| $) for the  A$ _{1g}$ mode is least,  indicating smallest change of this mode frequency (compared to E$ ^{1}_{2g} $ and B$ _{2g} $ ) with doping. Although DFT analysis qualitatively captures the experimental trend of  $\Delta\omega$ versus $ p $ for E$ ^{1}_{2g} $ and B$ _{2g} $  (Figs. \ref{2}, \ref{3}(c) and \ref{3}(d)), the relatively smaller phonon hardening observed in experiments for the A$ _{1g} $ mode ( as compared to E$ ^{1}_{2g} $ and B$ _{2g} $ modes) is not captured in our DFT analysis.

	In order to understand why  A$ _{1g} $ and B$ _{2g} $ are affected differently, we have  calculated the EPC of pristine bilayer without inclusion of SOC to understand these trends, as estimation of EPC with doping in FET geometry is not currently implemented in the QE code. The EPC of a mode \(\nu\) at momentum \textit{\textbf{q}} (with frequency \(\omega\)\textsubscript{\textit{\textbf{q}}\(\nu\)}) is calculated as \cite{PhysRevB.85.161403}
	\begin{widetext}
		\begin{equation}
			\lambda\textsubscript{\textbf{\textit{q}}\(\nu\)} =  \frac{2}{\hbar\omega\textsubscript{\textit{\textbf{q}}\(\nu\)}N(\epsilon\textsubscript{\textit{f}})}\sum_{k}\sum_{mn}|\tensor{g}{^q_k^\nu_+^,_q^i_,^j_k}|\textsuperscript{2}\times\delta(\epsilon\textsubscript{\textit{\textbf{k+q,i}}}-\epsilon\textsubscript{\textit{f}})\times\delta(\epsilon\textsubscript{\textit{\textbf{k,j}}}-\epsilon\textsubscript{\textit{f}}),
		\end{equation} 
	\end{widetext}where \(\omega\) and \textit{N}(\(\epsilon\)\textsubscript{\textit{f}}) are the phonon frequency and electronic density of states at the Fermi energy, respectively. The EPC matrix element is given by
	\begin{equation}
		\tensor{g}{^q_k^\nu_+^,_q^i_,^j_k} = (\frac{\hbar}{2M\omega\textsubscript{\textit{\textbf{q}}\(\nu\)}})^\frac{1}{2} <\psi\textsubscript{\textit{\textbf{k+q,i}}}|\Delta V\textsubscript{\textit{\textbf{q}}\(\nu\)}|\psi \textsubscript{\textit{\textbf{k},j}}>, 
	\end{equation} where \(\psi\)\textsubscript{\textit{\textbf{k},j}} is the electronic wave function with wave vector \textit{\textbf{k}} and energy eigenvalue \(\epsilon\)\textsubscript{\textit{k,j}} for band \textit{j}, and \textit{M} is the ionic mass. \(\Delta\)V\textsubscript{\textit{\textbf{q}}\(\nu\)} is the change in the self-consistent potential induced by atomic displacements of phonon \textit{\textbf{q}}\(\nu\).
	The calculated values of EPC of B$ _{2g} $ and A$ _{1g} $ modes are 0.016 and 0.011 respectively, consistent with the experimental observation that B$ _{2g} $ phonon is renormalized more with the hole doping.
	
	It is interesting to compare these trends with the phonon renormalization seen in n-doped monolayer MoS$_{2}$ \cite{PhysRevB.85.161403}. Electron doping in monolayer MoS$_{2}$ has contrasting effects on the frequencies of A$_{1g} $ and E$ ^{1}_{2g} $ optic modes \cite{PhysRevB.85.161403}. While A$_{1g} $ mode softens significantly ( $\sim$  7 cm$^{-1}$ at  $\sim$ 1.8$\times$10$^{13}$/cm$^{2}$), E$ ^{1}_{2g} $ remains unaffected \cite{PhysRevB.85.161403}. We can understand this contrast as follows:  monolayer MoS$_{2}$ is a direct band-gap semiconductor with a gap of $\sim$ 1.8 eV with the VBM and CBM at the K-point \cite{doi:10.1021/nl903868w}. The CBM at the K-point of MoS$_{2}$ has contribution from the out-of-plane d$_{z^2}$ orbital of Mo atoms \cite{PhysRevB.85.161403}. The A$_{1g}$ mode has the symmetry of the lattice, hence matrix element $<\psi\textsubscript{\textit{\textbf{k+q,i}}}|\Delta V\textsubscript{\textit{\textbf{q}}\(\nu\)}|\psi \textsubscript{\textit{\textbf{k},j}}>$ is non zero \cite{PhysRevB.85.161403}. In contrast, matrix element $<\psi\textsubscript{\textit{\textbf{k+q,i}}}|\Delta V\textsubscript{\textit{\textbf{q}}\(\nu\)}|\psi \textsubscript{\textit{\textbf{k},j}}>$ of in-plane vibrational mode E$ ^{1}_{2g} $ vanishes as it is orthogonal to A$_{1g}$ irreducible representation \cite{PhysRevB.85.161403}. In comparison, hole doping in bilayer 2H-MoTe$_{2}$ leads to occupation of states at the top of the valence band at the K-point, having dominance of in-plane d$_{xy}$ orbitals (odd symmetry states) of Mo. The crystal symmetry at K-point is point group C$_2$ which is a nontrivial subgroup of D\textsubscript{\textit{3d}} and the symmetry of the valence band is A\textsubscript{\textit{2u}}. The matrix element \(<\)\(\psi\)\textsubscript{\textit{\textbf{f}}}\(|\)\(\Delta\) V\textsubscript{\textit{\textbf{q}}\(\nu\)}\(|\)\(\psi\) \textsubscript{\textit{\textbf{i}}}\(>\) (where \textit{i} and \textit{f} are  the initial and final electronic wavefunctions) for \(\nu\) = A$ _{1g} $, E$ ^{1}_{2g} $ and B$ _{2g} $ modes are non-zero as calculated using direct product table for C$_2$. Hence, changes in occupancy of these states as a function of doping result in renormalization  of these modes. Though A$ _{1g} $ and B$_{2g}$ modes have different symmetries in bulk, the modes reduce to the same symmetry, A$ _{1g} $ in the case of bilayer (symmetry in even layer, odd layer, and bulk MoTe$ _{2} $ has been listed in table S1), softening is stronger for B$_{2g}$ mode as compared to A$ _{1g} $. This is consistent with the EPC  being slightly higher for the B$_{2g}$ (0.016) than the A$ _{1g} $ (0.011) mode and is also evident in frequency versus hole doping concentration plot (Fig.\ref{6}(e)).

	\section{Conclusions}
	
	In FET devices with  2, 3 and 7 layers of MoTe$ _{2} $ as channels, we have demonstrated that hole doping induces  phonon softening and linewidth broadening of the E$ ^{1}_{2g} $ and B$_{2g} $ modes, while the A$ _{1g} $ mode shows relatively small phonon hardening and linewidth sharpening. Due to dielectric screening, we find that holes are induced only in the top two layers of these channels upon electrochemical top gating, as evident in the layer dependent frequency softening of the E$ ^{1}_{2g} $ mode. Results of our first-principles density functional theory calculations agree qualitatively with our experiments. Interestingly, effects of EPC in hole doped MoTe$ _{2} $ observed here are in sharp contrast to the trends seen earlier in electron doped monolayer MoS$_{2}$. We explain this in terms  of the difference in  symmetry of their frontier states relevant to electron and hole doping. In addition to being relevant to use Raman spectroscopy as a non-invasive tool for characterization of MoTe$ _{2} $-FET devices, our study will be useful in understanding the role of relevant phonon interaction with charge carriers  in determining carrier mobility in MoTe$ _{2} $ devices.

	\section*{Acknowledgment}
	We thank the Centre for Nanoscience and Engineering department (CeNSE) of IISc for device fabrication facilities. AS is thankful to JNCASR for postdoctoral fellowship. AKS thanks Department of Science and Technology (DST), India for support under the Nanomission project and Year of Science Professorship. UVW acknowledges support from a J. C.  Bose National Fellowship of SERB-DST, Govt. of India and an AOARD project from United States Air Force. 	
	
	\bibliographystyle{apsrev4-2}
	\bibliography{ref}{}
	
\end{document}


\title{Supplemental Material\\Symmetry induced phonon renormalization in few layers of 2H-MoTe$ _{2} $ transistors: Raman and first-principles studies}

\author{Subhadip Das$^1$, Koyendrila Debnath$ ^{2} $, Biswanath Chakraborty$^{1,3}$, Anjali Singh$^{4}$, Shivani Grover$ ^{2} $, D. V. S. Muthu$^1$, U. V. Waghmare$^2$ and A. K. Sood$^1$}

\email{asood@iisc.ac.in}
\affiliation{$^1$Department of Physics, Indian Institute of Science, Bangalore 560012, India
	\\
	$^{2}$  Theoretical Sciences Unit, Jawaharlal Nehru Centre for Advanced Scientific Research, Bangalore-560064, India \linebreak $ ^{3} $ Present address: Department of Physics, Indian Institute of Technology Jammu, Jammu-181221, J\&K, India\\
	$ ^{4} $ Center for Study of Science, Technology \& Policy (CSTEP), Bangalore 560094, India.}

\maketitle
\vfill

\begin{table}[h]
\caption{Irreducible representation  of the Raman modes at $\Gamma$-point for N-layer and bulk MoTe$ _{2} $ \cite{doi:10.1021/acs.nanolett.5b02683_1}. The dagger symbols ($\dagger$) represent silent modes. The E$ _{1g} $ mode is absent in backscattering configuration \cite{doi:10.1021/acs.nanolett.5b02683_1}. The modes with E$ ^{'} $ symmetry are both Raman and infrared active \cite{doi:10.1021/acs.nanolett.5b02683_1}.}
\begin{center}
	\begin{tabular}{ ccccccc}
		\hline
	Layer number&\multicolumn{6}{c}{Zone-center phonon representation of the  Raman modes} \\
		\hline
		 &   $\leq$ 30 cm$ ^{-1} $ &  $\leq$ 40 cm$ ^{-1} $ &   $\sim$120 cm$ ^{-1} $  &   $\sim$ 170 cm$ ^{-1} $ &   $\sim$ 235 cm$ ^{-1} $ &   $\sim$ 290 cm$ ^{-1} $    \\
		Odd Layer& $\frac{N-1}{2} E'$  & $\frac{N-1}{2} A_{1}'$  & $\frac{N-1}{2} E'$  & $\frac{N+1}{2} A_{1}'$ & $\frac{N+1}{2} E'$ & $\frac{N-1}{2} A_{1}'$  \\ 
		Even layers & $\frac{N}{2} E_{g}$  & $\frac{N}{2} A_{1g}$  & $\frac{N}{2} E_{g}$  & $\frac{N}{2} A_{1g}$  & $\frac{N}{2} E_{g}$  & $\frac{N}{2} A_{1g}$   \\  
		Bulk & E$ _{2g} $ & B$ _{2g}^{\dagger} $ & E$ _{1g} $ & A$ _{1g} $ & E$ _{2g} $ & B$ _{2g} ^{\dagger}$\\    
		\hline 
	\end{tabular}

\end{center}

\end{table}
\clearpage
\section{E\lowercase{lectrical and} R\lowercase{aman spectral characterization of the bilayer} M\lowercase{o}T\lowercase{e}$ _{2} $ \lowercase{nanocrystal}}
\begin{figure}[ht!]	
	\includegraphics[width=0.95\textwidth]{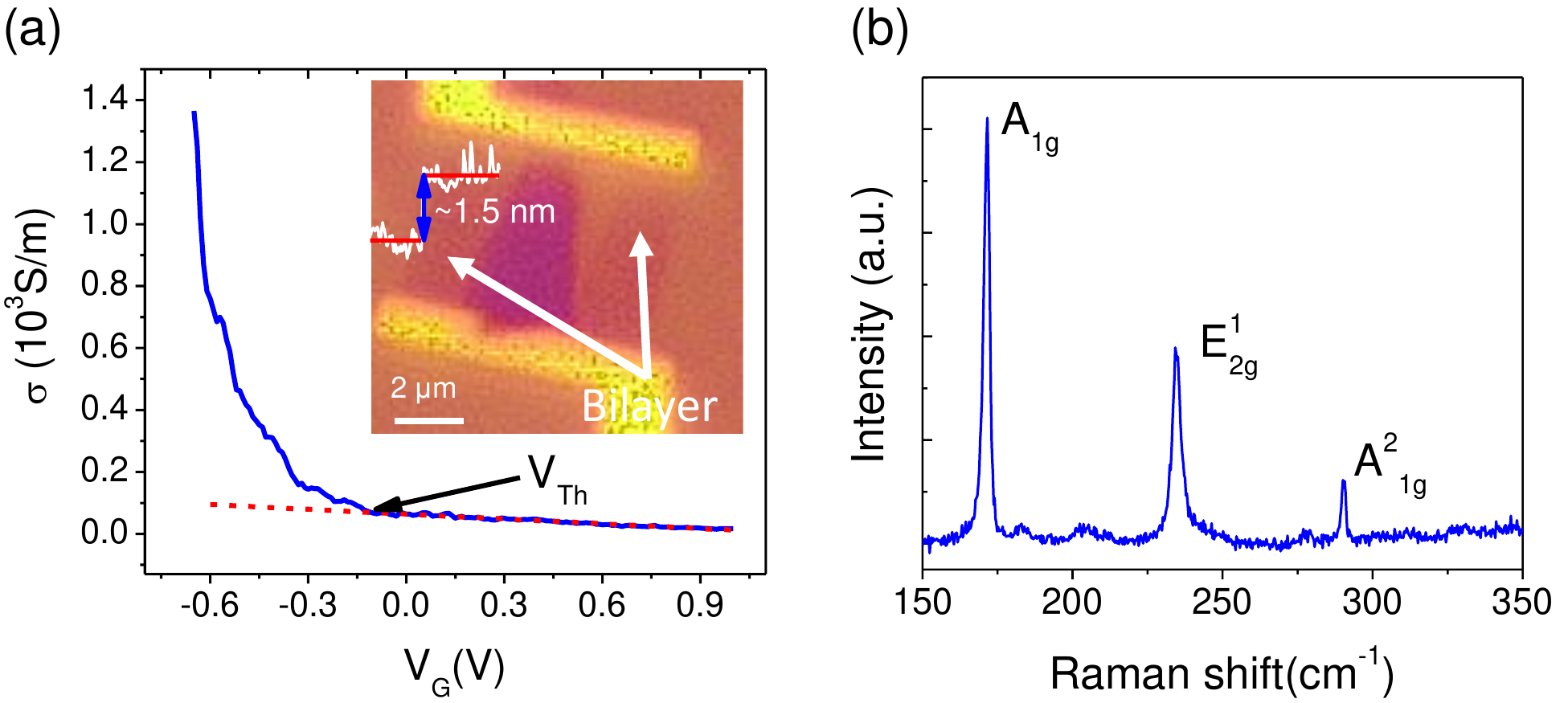}
	\caption{(a) Transport characteristics of the field-effect device with bilayer and multilayer ($\sim$ 9nm) parallel channel. The field effect mobility and current on/off ratio are $\sim$ 0.21 cm$ ^{2} $/V.s and 10$ ^{2} $, respectively. The red dashed line indicates linear fit to the transistor off state. The current threshold voltage ($\sim$-0.13V) is indicated in the figure. Inset shows the AFM height profile with  the optical image of the device. (b) Raman spectrum of the nanocrystal with 660 nm laser excitation.  }
	\label{s2b}
\end{figure}
\newpage
\section{R\lowercase{aman spectral background from the ionic liquid} }
\begin{figure}[ht!]	
	\includegraphics[width=0.9\textwidth]{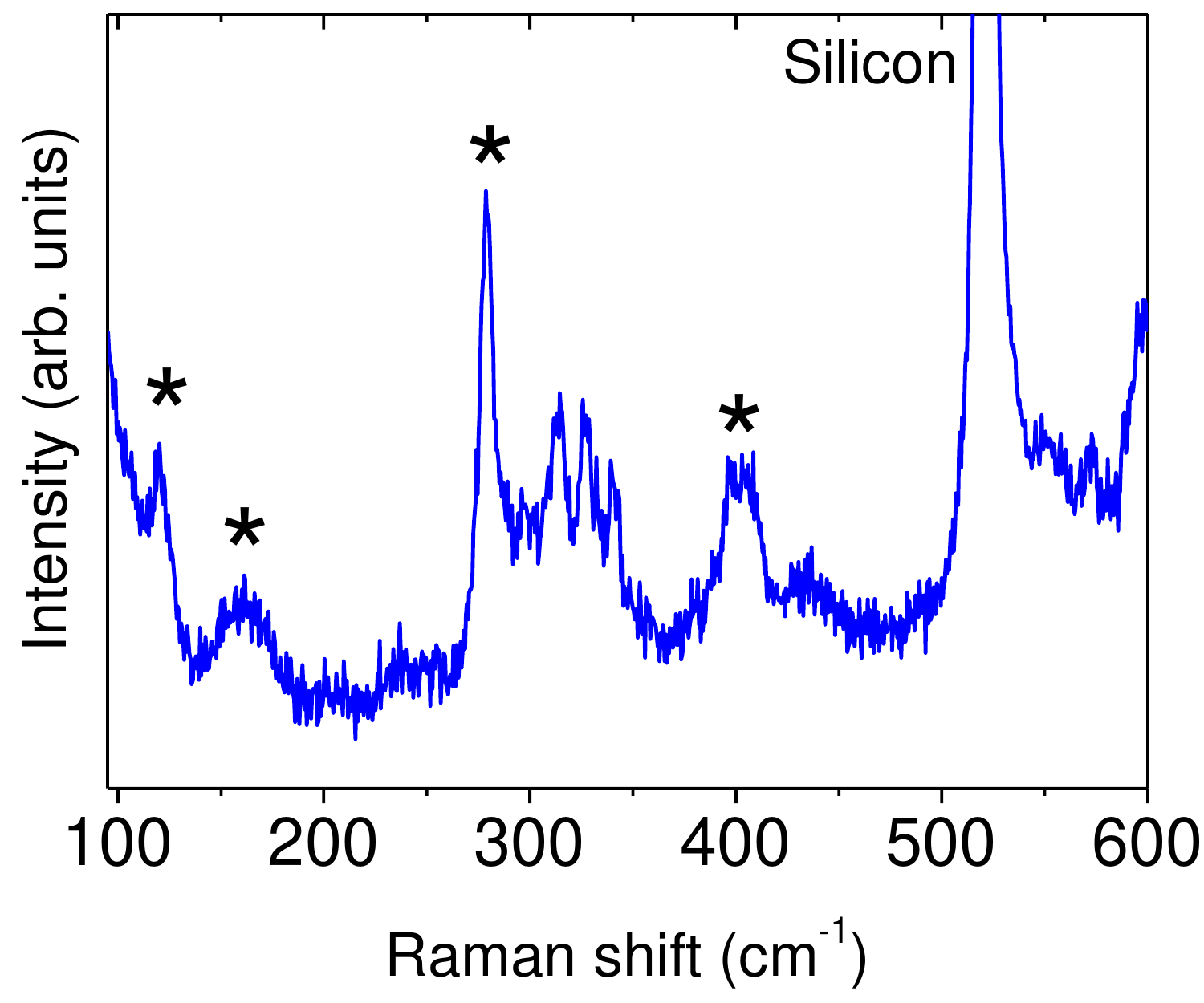}
	\caption{Raman spectrum of  EMIM-TFSI ionic liquid. The star symbol marks the Raman modes. The Raman mode at $\sim$ 520 cm$ ^{-1} $ is from the silicon substrate. }
	\label{s2b}
\end{figure}
\clearpage
\section{O\lowercase{utput characteristics of the devices at ambient} }
\begin{figure}[!htb]	
	\includegraphics[width=0.95\textwidth]{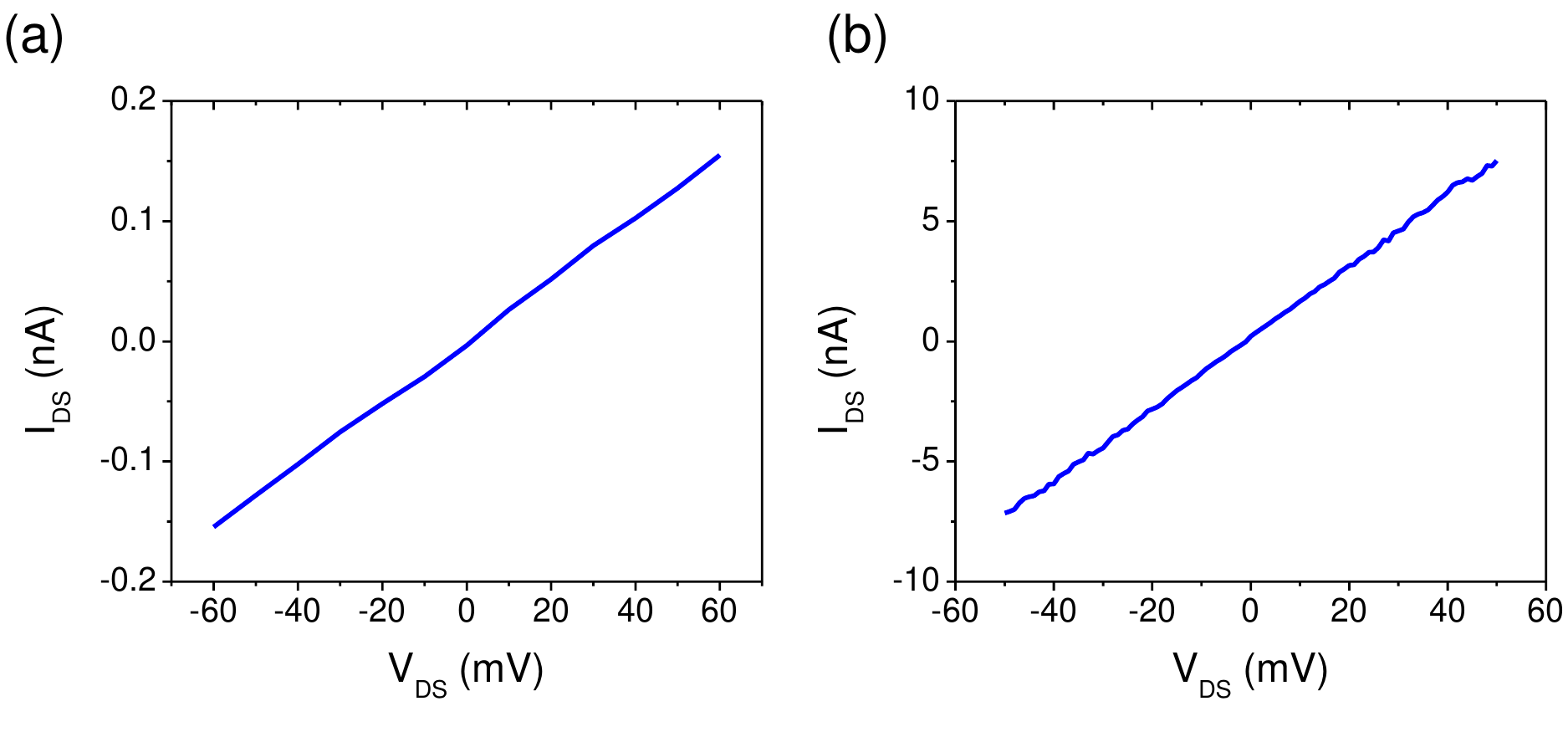}
		\caption{Linear drain current (I$ _{DS} $) versus drain voltage (V$ _{DS} $) measurement at zero gate bias for (a) three- and (b) seven-layer nanocrystals.   }
	\label{s2b}
\end{figure}

\bibliography{ref2}